\documentclass[fleqn,12pt,twoside]{article}
\usepackage[headings]{espcrc1}

\readRCS
$Id: espcrc1.tex,v 1.2 2004/02/24 11:22:11 spepping Exp $
\usepackage{graphicx}
\usepackage[figuresright]{rotating}

\newcommand{\AmS}{{\protect\the\textfont2
  A\kern-.1667em\lower.5ex\hbox{M}\kern-.125emS}}

\title{Thermodynamics of (2+1)-flavor QCD}

\author{C.~Schmidt\address[BNL]{Brookhaven National Laboratory,
        Upton, NY 11973, USA}
        and
        T.~Umeda\addressmark[BNL]
        for the RBC-Bielefeld Collaboration}
       
\runtitle{Thermodynamics of (2+1)-flavor QCD}
\runauthor{C.Schmidt and T.Umeda for the RBC-Bielefeld Collab.}

\begin{document}

\maketitle

\begin{abstract}
\vspace*{-4cm}\hfill{\tt BNL-NT-06/31}\\[3.5cm]

We report on the status of our QCD thermodynamics project. It is performed on
the QCDOC machine at Brookhaven National Laboratory and the APEnext machine at
Bielefeld University. Using a 2+1 flavor formulation of QCD at almost realistic
quark masses we calculated several thermodynamical quantities. In this
proceeding we show the susceptibilites of the chiral condensate and the
Polyakov loop, the static quark potential and the spatial string tension.  
\end{abstract}

\section{Introduction and Lattice Setup}

The calculation of QCD thermodynamics from first principle is
important for various research areas such as Heavy Ion Phenomenology,
Cosmology and Astrophysics. Lattice QCD enables us to carry out such
calculations. Especially for HIC phenomenology it is mandatory to improve
estimates on some basic thermodynamic quantities which have been obtained in
previous lattice calculations. Since thermodynamics of lattice QCD requires huge
computational resources, it is difficult to perform an ideal simulation.
Recent studies tell us that quark masses and the number of flavors strongly 
affect thermodynamic quantities \cite{Nf}. Reliable continuum extrapolations
are of tremendous importance as well \cite{a=0}.  
Therefore, it is our goal to study QCD thermodynamics with almost
realistic quark masses on the QCDOC machine at Brookhaven National
Laboratory and the APEnext machine at Bielefeld University.
The calculation is performed with $N_f=2+1$, which means 2 degenerate
light quarks and one heavier quark on lattices with $N_t=4$ and 6.
The lightest quark masses of our simulation yields a pion mass of about 150
MeV and kaon mass of about 500 MeV.  

For such calculations we adopt the p4fat3 quark action, which is
an improved Staggered quark action \cite{p4fat3}, 
with a tree-level improved Symanzik gauge action. 
By using the p4fat3 action,
the free quark dispersion relation has the continuum form
up to $O(p^4)$, and taste symmetry breaking is suppressed by a
3-link fattening term. 
The action also improves bulk thermodynamical quantities
in the high temperature limit \cite{p4fat3}.
The improvements are essential to control the continuum extrapolation 
on rather coarse lattices, i.e. $N_t=4$ and 6. 
The gauge ensembles are generated by an exact RHMC algorithm \cite{RHMC}.

As a status report of the project, in this proceeding, we present
several thermodynamics quantities, which are susceptibilities of the light and
strange quark chiral condensate, the Polyakov loop susceptibility, the static
quark potential, and the spatial string tension. The details of the critical
temperature calculation are given in our recent paper \cite{Tc}.

\section{Order Parameters and Susceptibilities}
To investigate the QCD critical temperature and phase diagram, order parameter
of the QCD transition are indispensable. 
In the chiral limit the chiral condensate $\langle
\bar{\psi}\psi \rangle$ is the order parameter for the spontaneous
chiral symmetry breaking of QCD. On the other hand in the heavy quark
limit the Polyakov loop $\langle L \rangle$ is the order parameter 
of the deconfinement phase transition.
For finite quark masses, these observables remain good
indicators for the (pseudo) critical point.
Especially their susceptibilities are useful to determine the critical
coupling $\beta_c$ in numerical simulations.

Figure~1 shows the susceptibilities of the chiral condensate and the
Polyakov loop. Their peak positions define the point of most drastic
change of each order parameters, i.e. the (pseudo) critical point of the QCD
transition. The results are interpolated in the coupling $\beta$ by using the 
multi-histogram re-weighting technique \cite{multi}.
\begin{figure}[t]
\begin{center}
\resizebox{140mm}{!}{
 \includegraphics{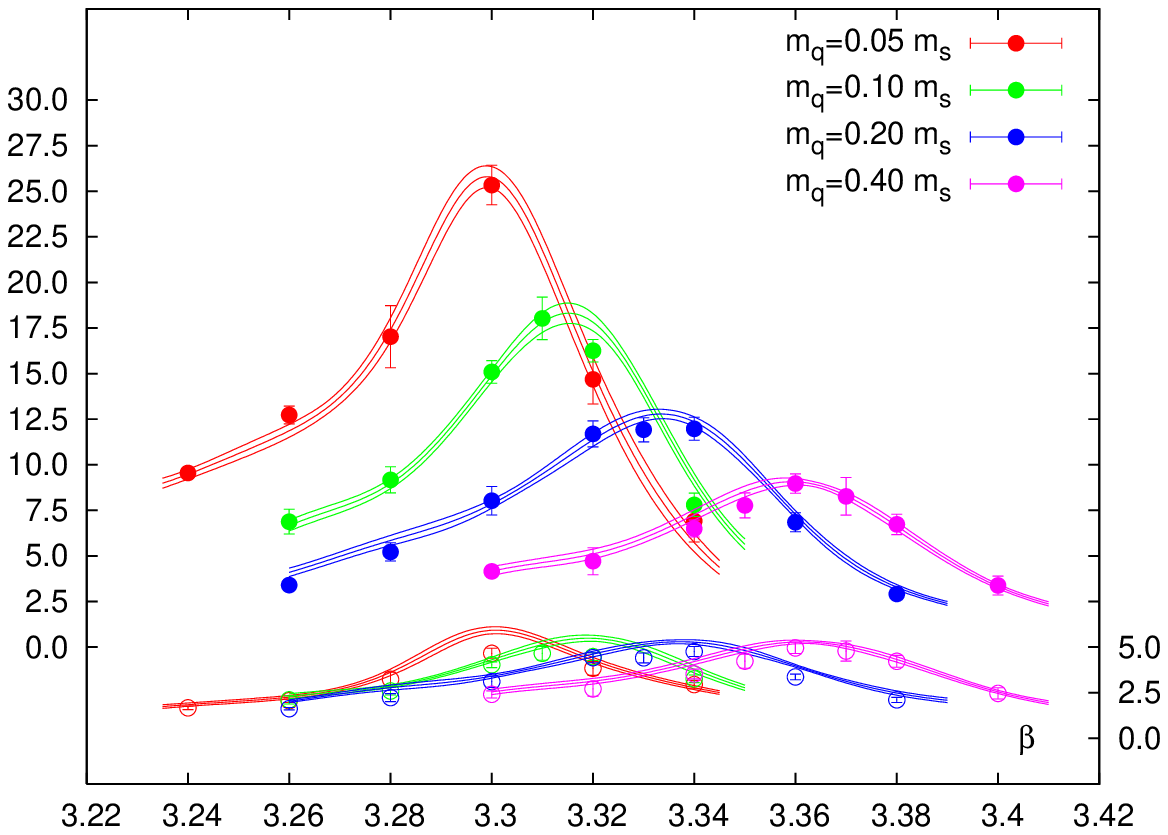}
 \includegraphics{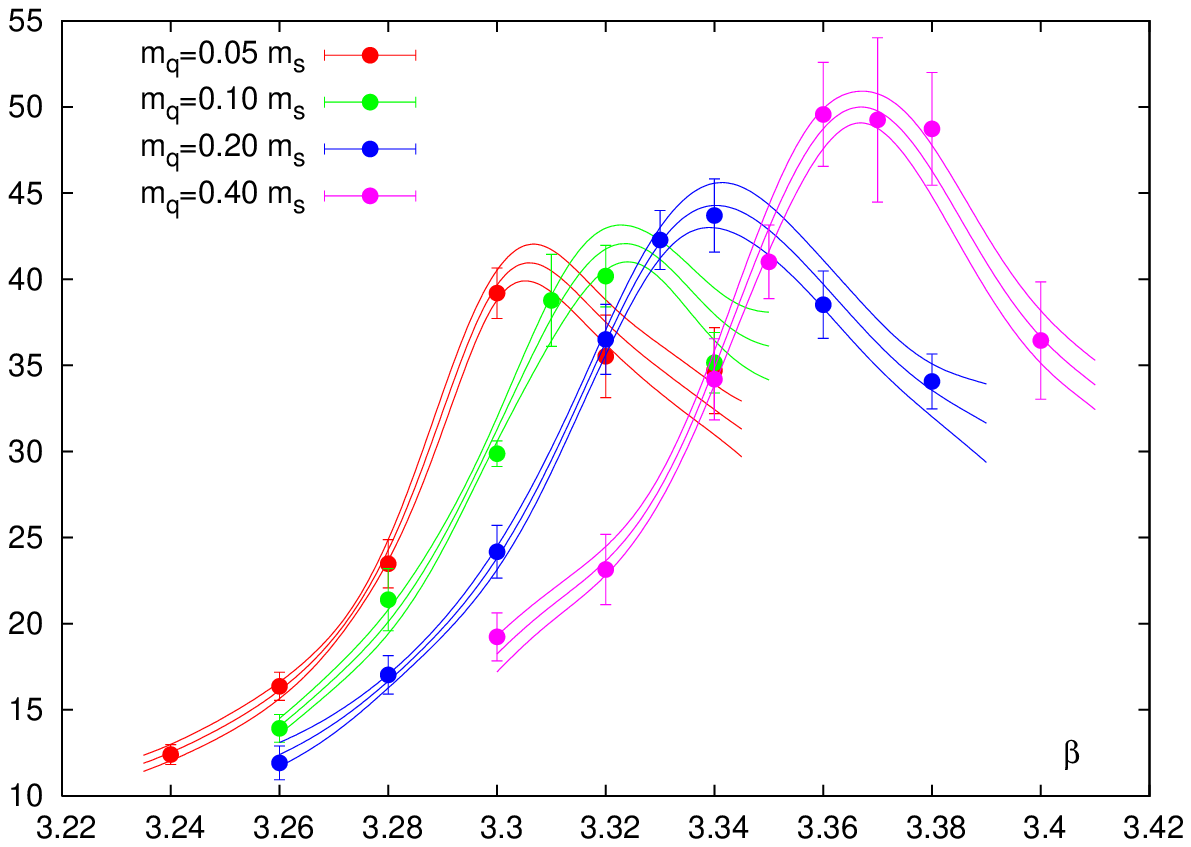}
}
\vspace*{-8mm}
\caption{The susceptibility of the chiral condensate (left) and the
 susceptibility of the Polyakov loop (right) on $8^3\times 4$ lattices.
In the left panel filled and open symbols are results for light and strange
 quarks, respectively. The lines are calculated by the multi-histogram
 re-weighting technique.}
\label{fig_suscep}
\end{center}
\end{figure}

The strength of the transition decreases with increasing quark masses,
this is reflected in the decreasing peak height of the chiral susceptibilities.
We calculate these susceptibilities on lattices with aspect ratios of
$N_s/N_t=2$ and 4. Since we see a rather small volume
dependence the results suggest that the transition is in fact not a true
phase transition in the thermodynamic sense but a rapid crossover. 
The peak position of the chiral and Polyakov loop susceptibilities are almost
identical, i.e. the chiral and deconfinement transition occur at almost
the same temperature. The discrepancy between the peak positions shrinks with
increasing volume.

\section{Scale Setting and the Heavy Quark Potential}
The lattice scale is determined from the heavy quark potential $V(r)$
which is extracted from Wilson loops.
The Wilson loop expectation values are calculated on $16^3 \times 32$ 
lattices with APE smearing in spatial direction.
The spatial path in a loop is determined by the Bresenham algorithm \cite{BA}.
We calculate the string tension, $\sigma$
and Sommer scale $r_0$, which is defined \cite{r0} as the distance where the 
corresponding force of the static quark potential
matches a certain value suggested by phenomenology:
$r^2\frac{\partial V}{\partial r}\mid_{r=r_0}=1.65$. 
To remove short range lattice artifacts we use the improved
distance, $r_{imp}$, which is defined as
\begin{equation}
\frac{1}{4\pi r_{imp}}\equiv \int\frac{d^3k}{(2\pi)^3}
\frac{e^{ikr}}{4\sum_i(\sin^2{\frac{k_i}{2}})
+\frac{1}{3}\sin^4{\frac{k_i}{2}}}.
\end{equation}
\begin{figure}[t]
\begin{center}
\resizebox{140mm}{!}{
 \includegraphics{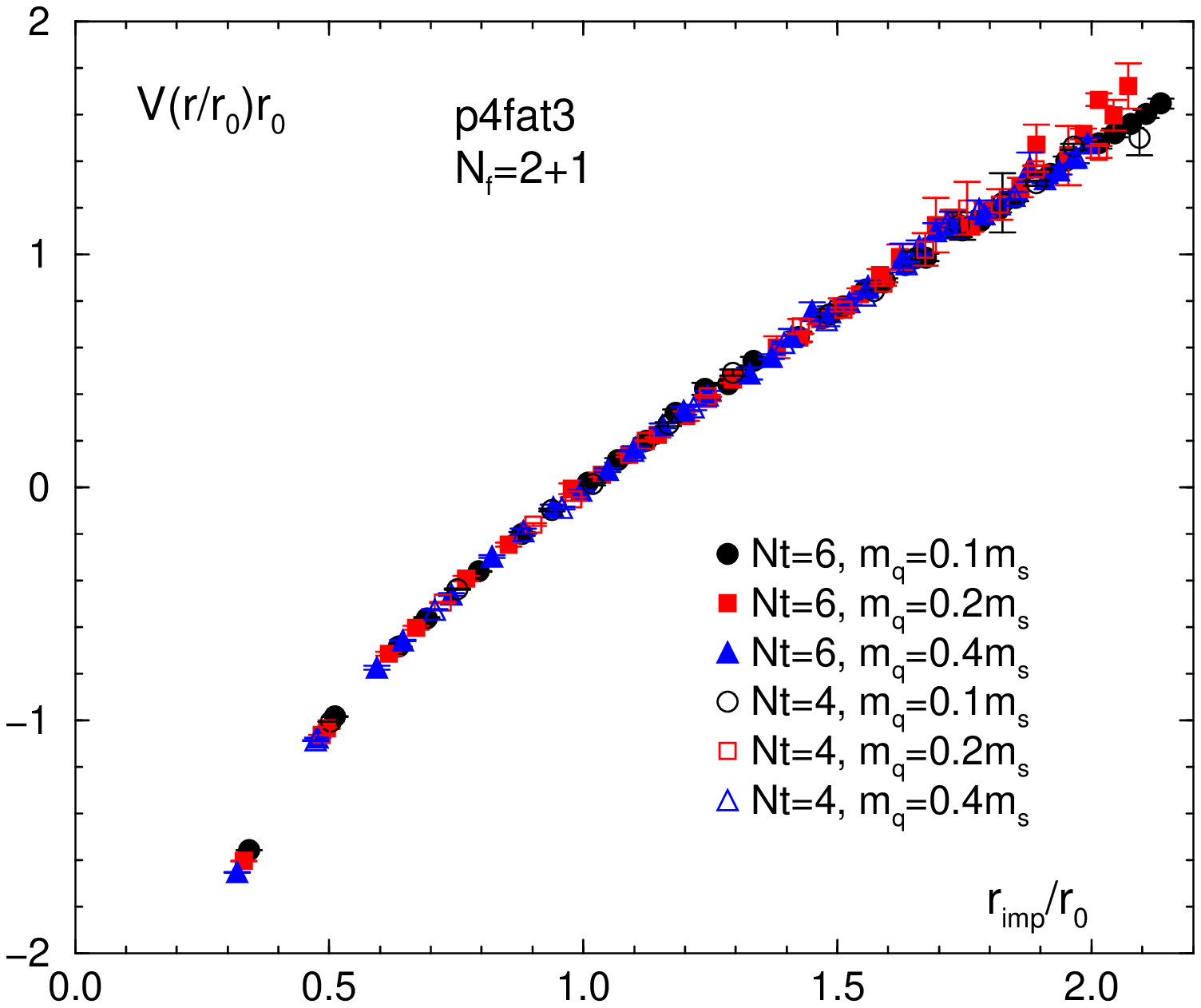}
\hspace{30mm} \includegraphics{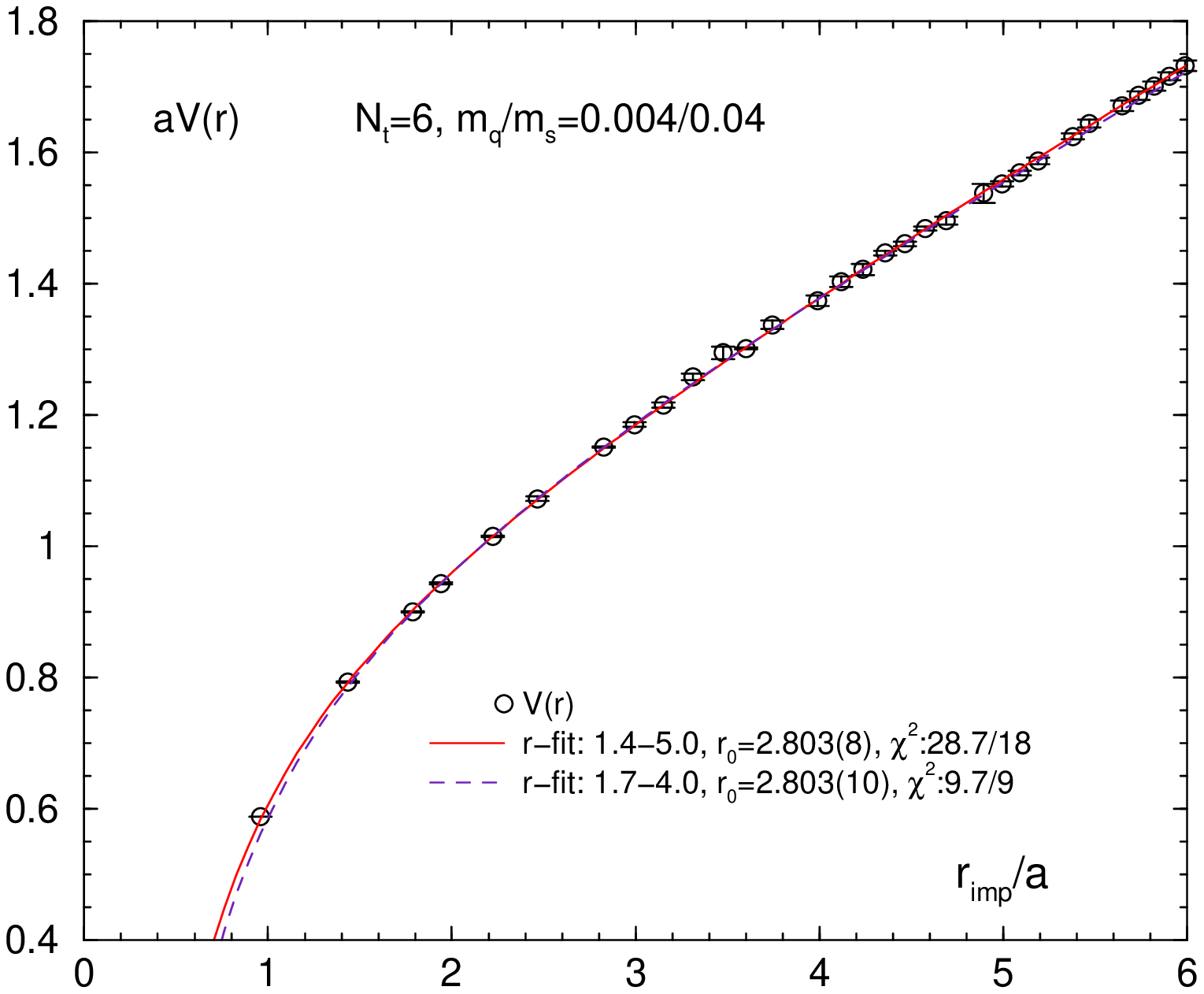}
}
\vspace*{-8mm}
\caption{Heavy quark potentials scaled by $r_0$ for various cut-offs and 
quark masses (left) and the fit-range dependence of $r_0$ (right). The latter shows
the result for $N_t=4, m_q/m_s=0.026/0.065$ and fits with the 3-params. fit-form.}
\end{center}
\end{figure}

In our lattice setup, we find almost no mass and cutoff dependence in
the potential scaled by $r_0$ at $N_\tau=4$ and 6 (Fig.2(left)). 
As discussed in previous studies \cite{sb}, 
we also find no string breaking effects even at large $r$.

To estimate systematic uncertainties of the potential fit, we performed 
several types of fits, e.g. different fit-ranges in $r$ (see Fig.2(right))
and fit-forms (3 \& 4 params. fits),
\begin{equation}
V(r)=C+\frac{\alpha}{r_{imp}}+\sigma r_{imp},
~~~
V(r)=C+\frac{\alpha}{r}+\sigma r+d(\frac{\alpha}{r_{imp}}-\frac{\alpha}{r}).
\end{equation}
The differences in the mean values of the fits are evaluated as a systematic
uncertainty of the scale setting. From a combined quark mass and cut off
extrapolation of $r_0T_c\equiv(r_0/a)(aT_c)$ we finally obtain a critical
temperature of $T_c=192(7)(4)$~MeV at the physical point \cite{Tc}. Here we
used $r_0=0.469$~fm to set the scale. The first error
summarizes all statistical and systematic errors on $r_0$ and the critical
couplings $\beta_c$ and the second error reflects the remaining uncertainties in 
the extrapolation.

\section{Spatial String Tension}

Let us now discuss the calculation of the spatial string tension which is
important to verify the theoretical concept of dimensional reduction at high
temperatures. The spatial string tension is extracted from the spatial static quark
``potential''  (from spatial Wilson loops).
We use the same analysis technique as for the usual (temporal) static quark
potential.

At high temperature, the spatial string tension $\sigma_s(T)$ is
expected to behave like 
\begin{eqnarray}
 \sqrt{\sigma_s(T)}&=&cg^2(T)T.\label{2loop}
\end{eqnarray}
Here $g^2(T)$ is the temperature dependent coupling constant from the 
2-loop RG equation,
\begin{eqnarray}
g^{-2}(T)&=&2b_0\ln{\frac{T}{\Lambda_\sigma}}+\frac{b_1}{b_0}
\ln{\left(2\ln{\frac{T}{\Lambda_\sigma}}\right)}.
\end{eqnarray}
If dimensional reduction works, the parameter``$c$'' should be equal to the
3-dimensional string tension and should be flavor independent.
\begin{figure}[t]
\begin{center}
\resizebox{80mm}{!}{
 \includegraphics{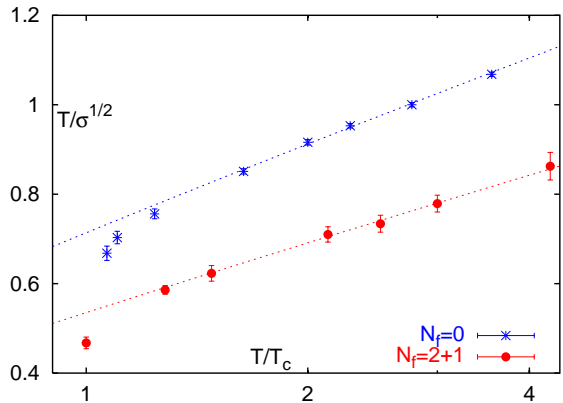}
}
\vspace*{-8mm}
\caption{Temperature dependence of the spatial string tension for
 $N_f=0$ \cite{spst} and $N_f=2+1$. Both dotted lines are fits with
 Eq.~\ref{2loop}. Note, that the scale on the horizontal axis is logarithmic.}
\label{sigma_s}
\end{center}
\end{figure}

Our 2+1 flavor result yields $c=0.587(41)$ and $\Lambda_\sigma/T_c=0.114(27)$,
obtained by a fit with Eq.~\ref{2loop}. On the other hand, we plot in
Fig.\ref{sigma_s} also the quenched result \cite{spst} which gives
$c=0.566(13)$ and $\Lambda_\sigma/T_c=0.104(9)$. We thus find that the 
parameter ``c'' is -- within statistical errors -- independent on the number
of dynamical flavors and that dimensional reduction works well even for
$T=2T_c$. This analysis can and will be refined in the future by taking into
account higher order corrections to Eq.~\ref{2loop}~\cite{Schroder:2005zd}.



\begin{thebibliography}{9}
\bibitem{Nf} F.~Karsch, E.~Laermann and A.~Peikert,
Nucl.\ Phys.\ B {\bf 605}, 579 (2001).
\bibitem{a=0} A.~Ali Khan {\it et al.}  [CP-PACS collaboration],
Phys.\ Rev.\ D {\bf 64}, 074510 (2001).
\bibitem{p4fat3} U.~M.~Heller, F.~Karsch and B.~Sturm,
Phys.\ Rev.\ D {\bf 60}, 114502 (1999).
\bibitem{RHMC} M.~A.~Clark, A.~D.~Kennedy and Z.~Sroczynski,
Nucl.\ Phys.\ Proc.\ Suppl.\  {\bf 140}, 835 (2005).
\bibitem{Tc} M.~Cheng {\it et al.},
arXiv:hep-lat/0608013, to appear in Phys.\ Rev.\ D.
\bibitem{multi} A.~M.~Ferrenberg and R.~H.~Swendsen,
Phys.\ Rev.\ Lett.\  {\bf 61} (1988) 2635.
\bibitem{r0} M.~Guagnelli, R.~Sommer and H.~Wittig,
Nucl.\ Phys.\ B {\bf 535}, 389 (1998).
\bibitem{BA} B.~Bolder {\it et al.},
Phys.\ Rev.\ D {\bf 63}, 074504 (2001).
\bibitem{sb} S.~Aoki {\it et al.}  [CP-PACS Collaboration],
Phys.\ Rev.\ D {\bf 60}, 114508 (1999).
\bibitem{spst} G.~Boyd {\it et al.}, 
Nucl.\ Phys.\ B {\bf 469}, 419 (1996).
\bibitem{Schroder:2005zd}
  Y.~Schr\"oder and M.~Laine,
  PoS {\bf LAT2005}, 180 (2006)
  [arXiv:hep-lat/0509104].
\end{thebibliography}
\end{document}